\documentstyle[prd,aps,twocolumn]{revtex}
\input{psfig}
\begin{document}
\draft
\fnsymbol{footnote}

\def\rhomin{\rho{}_{\scriptstyle -}}
\def\rhoplus{\rho{}_{\scriptstyle +}}
\def\umin{u{}_{-}}
\def\uplus{u{}_{+}}
\def\vmin{v{}_{-}}
\def\vplus{v{}_{+}}
\def\vdown{v{}_{\downarrow}}
\def\vup{v{}_{\uparrow}}
\def\udown{u{}_{\downarrow}}
\def\uup{u{}_{\uparrow}}

\wideabs{

\title{Phase Transitions and the Mass-Radius Curves of Relativistic Stars 
\footnotemark
}

\author{Lee Lindblom}
\address{Theoretical Astrophysics 130-33,
         California Institute of Technology,
         Pasadena, CA 91125}

\date{27 February 1998}
\maketitle

\begin{abstract} The properties of the mass-radius curves of
relativistic stellar models constructed from an equation of state with
a first-order phase transition are examined.  It is shown that the
slope of the mass-radius curve is continuous unless the discontinuity
in the density at the phase transition point has a certain
special value.  The curve has a cusp if the discontinuity is
larger than this value.  The curvature of the mass-radius curve
becomes singular at the point where the high density phase material
first appears.  This singularity makes the mass-radius curve appear
on large scales to have a discontinuity in its slope at this point,
even though the slope is in fact continuous on microscopic scales.
Analytical formulae describing the behavior of these curves are found
for the simple case of models with two-zone uniform-density equations
of state.

\end{abstract}

}

\narrowtext

\section{Introduction}
\label{I} 

The structures of spherically symmetric stellar models are usually
described in general relativity theory in terms of the functions
$\rho(r)$, $p(r)$, and $m(r)$: the total energy density, the pressure,
and the ``mass'' contained within a sphere of radius $r$.  These
functions satisfy Einstein's equation for a static spherically symmetric
spacetime with fluid source, which may be reduced to the following pair of
ordinary differential equations \cite{opp-volkoff}:
\footnote{Submitted to Physical Review D.}
\begin{equation}
\label{1.1}
{dp\over dr} = -(\rho+p) {m+4\pi r^3 p\over r(r-2m)},
\end{equation}

\begin{equation}
\label{1.2}
{dm\over dr}=4\pi r^2 \rho.
\end{equation}

\noindent Consider the families of solutions to these equations, each
of which is determined by a different equation of state
$\rho=\rho(p)$.  For each equation of state and for each value of the
central pressure, there exists a unique solution of
Eqs.~(\ref{1.1})--(\ref{1.2}) that is non-singular at the center of
the star \cite{schmidt-rendall}.  Thus for each equation of state
there exists a one-parameter family of stellar models parameterized by
$p_c$ the central pressure of the star.  A large class of equations of
state have stellar models with finite total radii, $p(R)=0$, and
finite total masses $M=m(R)$.  The discussion here is limited to these
equations of state \cite{note1}.  The collection of total masses
$M(p_c)$ and radii $R(p_c)$ for a given equation of state is called
the mass-radius curve: $[M(p_c),R(p_c)]$.  Each equation of state
determines a unique mass-radius curve, and conversely it appears
(although the argument \cite{lindblom} falls short of being a rigorous
proof) that each mass-radius curve determines a unique equation of
state.  Thus there is hope that the high density equation of state of
neutron star matter may one day be determined by measurements of the
macroscopic mass-radius curve of these stars.

This paper is concerned with analyzing the features of the mass-radius
curve for the case of an equation of state with a first-order phase
transition.  Such a phase transition may well be a feature of the
equation of state of real neutron star matter.  Pion condensation
\cite{pionrefs} and/or quark deconfinement \cite{quarkrefs} might well
provide the mechanism that drives such a phase transition. This paper
does not focus on the microphysics of the mechanism that may trigger
such a transition, but rather the consequences that such a transition
might have on the observable macroscopic equilibrium structures of
neutron stars.  In particular this paper investigates how the
properties of such a phase transition might be read from the structure
of the mass-radius curve.

Consider equations of state that are smooth except at one value of the
pressure $p_{t}$ where the energy density undergoes a simple
discontinuity

\begin{equation}
\label{1.3}
\rhomin\equiv\lim_{p\uparrow p_{t}}\rho(p) <
\lim_{p\downarrow p_{t}}\rho(p) \equiv \rhoplus
\end{equation}

\begin{figure}
\centerline{\psfig{file=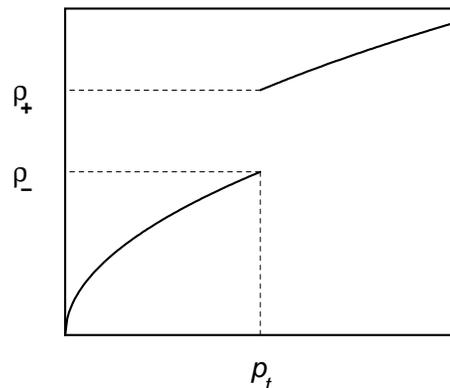,height=2.0in}}
\caption{Equation of state with a first-order phase transition
at the pressure $p=p_{t}$.}
\label{fig1}
\end{figure}

\noindent as illustrated in Fig.~\ref{fig1}.  It is convenient to
parameterize the magnitude of the discontinuity in the equation of state
by the dimensionless quantity $\Delta$:

\begin{equation}
\label{1.4}
\Delta = {\rhoplus-\rhomin\over\rhomin+p_{t}}.
\end{equation}

\begin{figure}
\centerline{\psfig{file=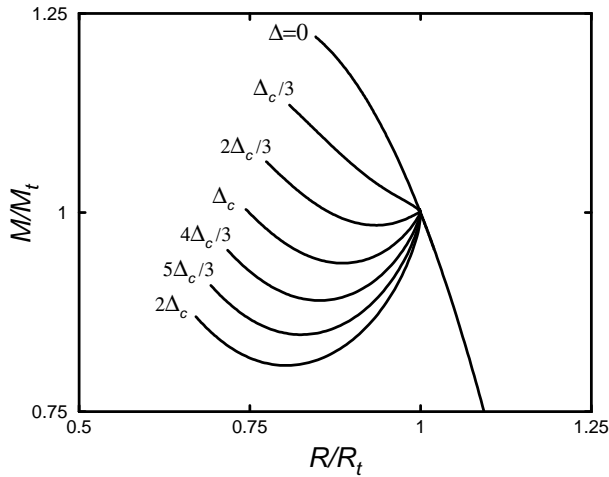,height=2.5in}}
\vskip .2cm
\caption{Mass-Radius curves for equations of state with first
order phase transitions.  The different curves correspond to
different values of the parameter $\Delta=(\rhoplus-\rhomin)/(\rhomin+p_{t})$.}
\label{fig2}
\end{figure}

\noindent Figure \ref{fig2} illustrates the mass-radius curves for the
equations of state shown in Fig.~\ref{fig1} (simple polytropes
$p\propto \rho^2$ with a density discontinuity inserted at 
the pressure $p_{t}$).  The mass scale, $M_{t}$, and radius
scale, $R_{t}$, used here are the total mass and radius of the stellar
model with central pressure $p_{t}$.  The quantity $\Delta_c$ that is
used to scale the discontinuity in the equation of state is defined by

\begin{equation}
\Delta_c = {\rhomin+3p_{t}\over 2(\rhomin+p_{t})}.
\label{1.5}
\end{equation}

\noindent (Note that $1/2\leq\Delta_c<3/2$.)  Figure~\ref{fig2}
illustrates that a first-order phase transition makes the mass-radius
curve bend sharply at the critical point $(M_{t},R_{t})$ where the
high density phase material first appears in the core of the star.

\begin{figure} \centerline{\psfig{file=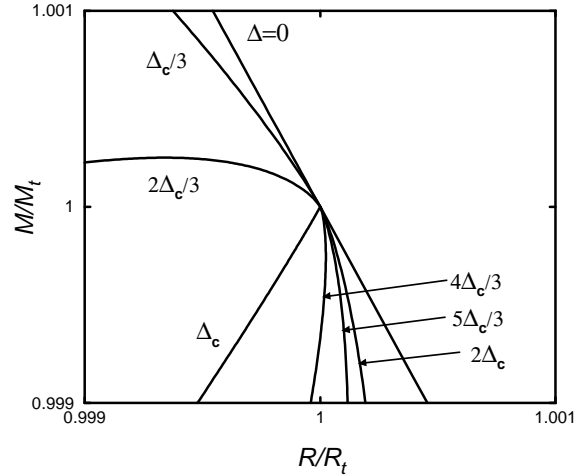,height=2.5in}} \vskip
.2cm \caption{Mass-Radius curves for equations of state with first order
phase transitions.  This figure represents the same stellar models
depicted in Fig.~\ref{fig2} but on a much finer scale.} \label{fig3}
\end{figure}

Figure~\ref{fig2} makes it appear that mass-radius curves have finite
discontinuities in their slopes at the point where the higher density
phase material first enters the star.  Further, it appears that the
magnitude of the discontinuity in the slope is determined by the
parameter $\Delta$ that measures the magnitude of the phase
transition.  Thus one might hope that an expression can be derived
which determines the properties of the phase transition (e.g.  the
value of $\Delta$) in terms of some features (e.g.  the change in
slope) of the mass-radius curve in a neighborhood of the point
$(M_{t},R_{t})$.  This hope is diminished, however, on closer
examination of these curves.  Figure~\ref{fig3} illustrates the same
set of mass-radius curves as shown in Fig.~\ref{fig2}, however, on a
much finer scale.  Figure~\ref{fig3} shows that the slopes of all of
the curves are in fact continuous at the point $(M_{t},R_{t})$, except
for the special case with $\Delta= \Delta_c$.  The mass-radius curves
for equations of state with $\Delta>\Delta_c$ reverse direction at the
point $(M_{t},R_{t})$, however their slopes are continuous there.  The
curves of models with strong first-order phase transitions have cusps
at the critical point.  This microscopic continuity of the slope makes
it impossible to find a purely local relationship between the
properties of the phase transition and the magnitude of the
macroscopic bend that occurs in the mass-radius curves, as illustrated
in Fig.~\ref{fig2}.

The continuity of the slope of the mass-radius curve even in the
presence of first-order phase transitions was first discovered in the
context of Newtonian stellar models by Ramsey~\cite{ramsey} and
Lighthill~\cite{lighthill}.  The corresponding result for relativistic
models was demonstrated in analogy with the Newtonian analysis by
Seidov~\cite{seidov}.  A more complete and somewhat more rigorous
derivation of this fact is presented in the Appendix here for
arbitrary relativistic stellar models.  These analyses demonstrate
that a special value of the magnitude of the phase transition
is $\Delta=\Delta_c$, with $\Delta_c$ defined in Eq.~(\ref{1.5}).  For
stronger phase transitions, $\Delta>\Delta_c$, the mass-radius curve
reverses direction at the critical point, and probably triggers the
onset of instability in the stellar models immediately above this
point.

The general analysis of Lighthill and Seidov shows that the slope of
the mass-radius curve is continuous (in almost all cases) even at the
critical stellar model where the influence of a phase transition is
first felt.  This result, however, raises more questions than it
answers.  The ``typical'' mass-radius curves displayed in
Figs.~\ref{fig2}--\ref{fig3} show that the phase transition does have
a very profound effect on the slope of these curves in a very small
neighborhood of the critical stellar model.  How does the phase
transition change the curvature of these curves on very small scales
in a neighborhood of the critical model, while leaving it relatively
unaffected on larger scales?  In Section \ref{sectII} a more detailed
analysis of the structure of the mass-radius curve in the neighborhood of
a critical point is undertaken in an attempt to understand this
behavior.  An analysis is given there of the simple case of equations
of state having two uniform-density zones: $\rho=\rhomin$ for
$p<p_{t}$ and $\rho=\rhoplus$ for $p>p_{t}$.  Analytical expressions
are derived for the mass-radius curve for these models in a small
neighborhood of the critical point $(M_{t}, R_{t})$.  These
expressions show that the phase transition causes the curvature of the
mass-radius curve to diverge at this point, even though its slope is
well defined and continuous there.  This singular part of the curvature
causes the mass-radius curves in these simple models to bend on
relatively small scales, much like the more realistic ones depicted in
Figs.~\ref{fig2}--\ref{fig3}.  Unfortunately, the two-zone models are
too simple to model accurately the behaviors of the mass-radius curves
of more realistic equations of state.  A more complicated analysis is
needed, but that analysis is deferred to a future investigation.


\section{Two-Zone Uniform-Density Models} \label{sectII}

The general solution of Einstein's equation representing a static
spherical uniform-density star was first found by
Schwarzschild~\cite{schwarzschild}.  Let $\rho_i$ denote the constant
density of the star.  Then the general solution to
Eqs.~(\ref{1.1})--(\ref{1.2}) can be written:

\begin{equation}
m(r)={4\pi\over 3}\rho_i r^3 + a_i,
\label{2.1}
\end{equation}

\begin{equation}
p(r)+\rho_i = b_i f_i(r)
\biggl[1+4\pi b_i\int^r_{c_i} r'f_i^3(r')dr'\biggr]^{-1},
\label{2.2}
\end{equation}

\noindent where $a_i$, $b_i$ and $c_i$ are arbitrary constants, and
$f_i(r)$ is defined as

\begin{equation}
f_i(r)=\biggl(1-{8\pi\over 3}\rho_i r^2 - {2a_i\over r}\biggr)^{-1/2}.
\label{2.3}
\end{equation}

\noindent More complicated stellar models composed of concentric
uniform-density layers may also be constructed by combining together
the basic solutions given in Eqs.~(\ref{2.1})--(\ref{2.2}).  These
laminated models satisfy Eqs.~(\ref{2.1})--(\ref{2.2}) with $\rho_i$
the fluid density within a particular layer.  The regularity of the
global solution is assured by choosing $c_i$ to be the inner radius of
the $i{}^{\rm th}$ layer, and the constants $a_i$ and $b_i$ to make
$p(r)$ and $m(r)$ continuous at $r=c_i$.

Now consider the stellar models composed of material having
a simple two-zone uniform-density equation of state:

\begin{eqnarray}
\rho = \Big\{
       \begin{array}{c}
             \rhomin\quad {\rm for}\quad p<p_{t},\\
              \rhoplus\quad {\rm for}\quad p>p_{t}. 
       \end{array}
\label{2.4}
\end{eqnarray}

\noindent This is the simplest equation of state having a first order
phase transition.  The family of stellar models associated with this
equation of state is easily obtained from Eqs.~(\ref{2.1})--(\ref{2.2}).

For stars with small central pressures $p_c < p_{t}$, the solutions
are the standard interior Schwarzschild
models~\cite{schwarzschild}.  These may be obtained from the
general expressions above by setting $a_i=c_i=0$ and
$b_i=p_c+\rhomin$.  The total masses and radii of these models are
determined by solving Eqs.~(\ref{2.1})--(\ref{2.2}) for the points where
$p(R)=0$ and $M=m(R)$.  As functions of the central pressure these
solutions are:

\begin{equation}
R^2(p_c)={3\over 8\pi\rhomin}\left[
1-\left({\rhomin+p_c\over\rhomin+3p_c}\right)^2\right],
\label{2.5}
\end{equation}   

\begin{equation}
M(p_c)={4\pi \over 3} \rhomin R^3(p_c).
\label{2.6}
\end{equation}

\noindent Thus the models with small central pressures have the
well-known cubic mass-radius curve.  The critical model in this family
is the one having $p_c=p_t$ with total mass $M_t = M(p_t)$ and total
radius $R_t=R(p_t)$.  Expressions for these critical values are given
by Eqs.~(\ref{2.5})--(\ref{2.6}) with $p_c=p_{t}$.

The stars with large central pressures, $p_c > p_{t}$, have
two concentric layers.  The inner layer is composed of high density
material with $\rho=\rhoplus$, and the outer layer of lower density
material with $\rho=\rhomin$.  The structure of the inner core is
determined from Eqs.~(\ref{2.1})--(\ref{2.2}) with $\rho_i=\rhoplus$,
$a_i=c_i=0$, and $b_i=p_c+\rhoplus$.  The radius, $r_t$, of this inner
core is determined by solving Eq.~(\ref{2.2}) for the point where
$p_{t}=p(r_t)$.  As a function of the central pressure, $p_c$, this
core radius is

\begin{equation}
r^2_t={3\over 8\pi\rhoplus}\left\{
1-\left[{(\rhoplus+p_c)(\rhoplus+3p_{t})
\over(\rhoplus+p_{t})(\rhoplus+3p_c)}\right]^2\right\}.
\label{2.7}
\end{equation}   

\noindent The outer envelopes of these models are determined again by
Eqs.~(\ref{2.1})--(\ref{2.2}) with $\rho_i=\rhomin$,
$a_i=4\pi(\rhoplus-\rhomin)r^3_t/3$,
$b_i=(\rhomin+p_{t})\sqrt{1-8\pi\rhoplus r_t^2/3}$, and $c_i=r_t$.
The quadrature indicated in Eq.~(\ref{2.2}) can be expressed
in terms of standard elliptic integral functions for this case, but that
representation does not offer any particular insight for our purposes 
here.

\begin{figure} \centerline{\psfig{file=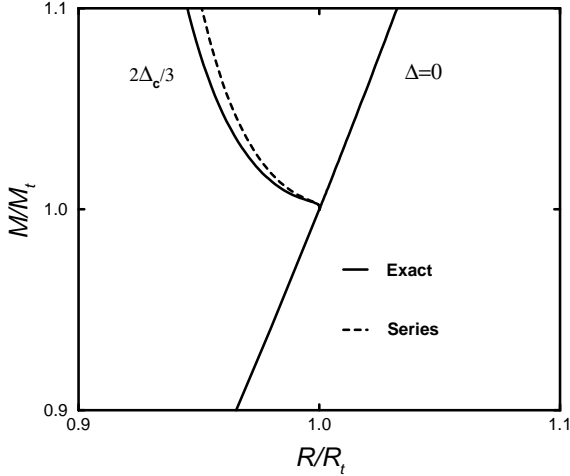,height=2.5in}} \vskip
.2cm \caption{Mass-Radius curve for the two-zone uniform-density
equation of state.  The series expansion for the curve (dashed line)
is compared to the exact (solid line) for a model with
$\Delta=2\Delta_c/3$ and $p_{t}/\rhomin=0.5$.} \label{fig4}
\end{figure}

The total masses and total radii of the two-zone uniform-density
models are found by solving Eqs.~(\ref{2.1})--(\ref{2.2}) for the
points where $p(R)=0$ and $M=m(R)$.  These equations can not be solved
analytically even for these simple two-zone models.  However, the
nature of the solutions near the critical model can be studied by
means of power series expansions.  The small parameter $s=(r_t/R_t)^2$
(which vanishes as $p_c\downarrow p_{t}$) can be used to expand the
various quantities (i.e., $b_i$ and $f_i$) that appear in
Eq.~(\ref{2.2}).  The integration that appears in Eq.~(\ref{2.2}) is
performed term by term and the resulting equation is solved for the
total radius of the star $p(R)=0$.  The resulting series expansion for
$R$ is used to evaluate the total mass $M=m(R)$ using Eq.~(\ref{2.1})
to the same order of approximation:

\begin{equation}
{R(s)\over R_t}=1+r_{1}s+r_{3/2}s^{3/2}+r_2s^2+{\cal O}(s^{5/2}),
\label{2.8}
\end{equation}

\begin{equation}
{M(s)\over M_t}=1+m_{1}s+m_{3/2}s^{3/2}+m_2s^2+{\cal O}(s^{5/2}),
\label{2.9}
\end{equation}

\noindent where the expansion coefficients $r_1$, $m_1$ etc. are given by

\begin{eqnarray}
r_1&=&{\Delta_c-\Delta\over 8\Delta_c^3},\label{2.10}\\
r_{3/2}&=&
         {\Delta(3-6\Delta_c^2-8\Delta_c^4)\over 8\Delta_c^4(3-2\Delta_c)}
          ,\label{2.11}\\
r_2&=&
{9\Delta(4\Delta_c-\Delta)(4\Delta_c^2-1)\over 128\Delta_c^6(3-2\Delta_c)}
     -{r_1^2\over 2},\label{2.12}\\
m_1&=&3r_1,\label{2.13}\\
m_{3/2}&=&
{\Delta(9-18\Delta_c^2-8\Delta_c^4)\over 8\Delta_c^4(3-2\Delta_c)}
          ,\label{2.14}\\
m_2&=&3(r_2+r_1^2),\label{2.15}
\end{eqnarray}

\noindent and where $\Delta$ and $\Delta_c$ are defined in
Eqs.~(\ref{1.4}) and (\ref{1.5}) respectively.  These series expansions
give a reasonably good approximation of the mass-radius curve near the
critical model as illustrate in Fig.~\ref{fig4}.  The series agree
with the exact (numerically determined) mass-radius curves to within
1\% for models whose masses and radii differ from the critical values
by up to about 10\% in the worst case examined.  (The series converge
most poorly for $\Delta=\Delta_c$ among those cases examined.)  This
level of accuracy is good enough to account for the interesting
features of the mass-radius curve near the critical point.
Figure~\ref{fig4} illustrates that the series exhibit the same
apparent discontinuity in slope on large scales as the exact curves.

The tangent vector to the mass-radius curves is determined by
differentiating the expressions for $x(s)\equiv R(s)/R_t$ and
$y(s)\equiv M(s)/M_t$ in Eqs.~(\ref{2.8})--(\ref{2.9}):

\begin{equation}
\left(\begin{array}{c} dx/ ds \\ dy/ ds \end{array}\right)
=\left(\begin{array}{c} r_1+ {3\over 2}r_{3/2}s^{1/2}+2r_2s
                  \\ m_1+ {3\over 2}m_{3/2}s^{1/2}+2m_2s\end{array}\right).
\label{2.17}
\end{equation}

\noindent This expression illustrates that the slope of the mass
radius curve is continuous even at the critical model.  This follows
from the fact that the tangent vector computed from just above the
critical model using Eq.~(\ref{2.17}) is $r_1(dx/dp_c)^{-1}$ times
that computed from just below using Eq.~(\ref{2.6}).  Just above the
critical point the tangent vector is proportional to the quantity
$r_1$ defined in Eq.~(\ref{2.10}).  $r_1$ is positive for weak phase
transitions $\Delta<\Delta_c$ but negative for strong transitions
$\Delta>\Delta_c$.  Thus the tangent vector and hence the mass-radius
curve itself reverses direction when a strong phase transition occurs.
This confirms the general continuity analysis of Ramsey, Lighthill,
and Seidov for the simple case of the two-zone uniform-density models.

While the general analysis of the structure of the mass-radius curve
fails for the case of a phase transition with $\Delta=\Delta_c$, the
analysis here for the simple two-zone uniform-density models succeeds
even in this case.  The terms proportional to $s$ in
Eqs.~(\ref{2.8})--(\ref{2.9}) vanish when $\Delta=\Delta_c$, and
therefore $s$ is not a good affine parameter for the mass-radius curve
at $s=0$ in this case.  Instead, the appropriate parameter is
$\lambda=s^{3/2}$.  In this case the tangent vector evaluated at the
critical model is $(dx/d\lambda,dy/d\lambda)=(r_{3/2},m_{3/2})$.  This
vector is {\it not} proportional to the tangent vector just below the
critical model, and so the slope of the mass-radius curve is {\it not}
continuous in this case.

In order to discuss the magnitude of the change in slope that occurs
at the critical model it is necessary to adopt a metric structure for
the space of masses and radii.  This makes it possible to define the
inner products of tangent vectors (and so define angles) and also more
generally to discuss the curvatures of these curves.  There is no
canonical choice for this metric, and therefore no absolute intrinsic
meaning can be given to angles or curvatures that are computed.
Nevertheless these quantities are useful tools for understanding the
features of the mass-radius curves seen in
Figs.~\ref{fig2}--\ref{fig4}.  Thus the metric used to display those
figures is adopted: the flat metric with Cartesian coordinates
$x\equiv R/R_t$ and $y\equiv M/M_t$.

Return now to the kink that occurs in the mass-radius curve for the
case of a phase transition with $\Delta=\Delta_c$.  The angle between
the slopes above and below the transition point can be determined
(using the metric defined above) from the inner product between the
tangent vectors.  The resulting angle $\theta$ depends solely and
monotonically on the ratio $p_t/\rhomin$:

\begin{equation}
\cos\theta={r_{3/2}+3m_{3/2}\over\sqrt{10(r_{3/2}^2+m_{3/2}^2)}}.
\label{2.16a}
\end{equation}

\noindent This $\theta$ varies from about $4.4^\circ$ for
$p_t/\rhomin=0$ to about $170.4^\circ$ for $p_t/\rhomin=\infty$.  This
formula is a simple example of the kind of relationship that one had
hoped to find relating the parameters of the phase transition and the
macroscopic structure of the mass-radius curve.  In this special case
(phase transitions with $\Delta=\Delta_c$) the magnitude of the kink
in the mass-radius curve determines the ratio $p_t/\rhomin$.
Unfortunately, this formula is not universal even for phase
transitions with $\Delta=\Delta_c$.  The magnitude of the kink
displayed in Fig.~\ref{fig3} does not satisfy this equation for
example.  The general form of this relationship must depend on other
features of the equation of state (e.g. $d\rho/dp$ at the transition
point) that are not present in the simple two-zone uniform-density
models.

The curvature of any of the mass-radius curves can be evaluated 
by differentiating the unit tangent vector along the trajectory
of the curve.  The resulting acceleration is equal to the inverse of
the radius of curvature of the curve.  For a general curve in a flat
two-dimensional space, this acceleration is given by

\begin{equation}
a=\left({d^2x\over ds^2}{dy\over ds}-{d^2y\over ds^2}{dx\over ds}\right)
\left[\left({dx\over ds}\right)^2+\left({dy\over ds}\right)^2\right]^{-3/2}.
\label{2.18}
\end{equation}

\noindent This expression is invariant under changes in the
parameterization of the curve, but not on the assumed metric of the
mass-radius space.  It is straightforward to evaluate this expression
using the series expansions, Eqs.~(\ref{2.8})--(\ref{2.9}), for the
curve:

\begin{eqnarray}
a&=&{3(3r_{3/2}-m_{3/2})\over40\sqrt{10}r_1^2}
\left[{1\over s^{1/2}}-{9(r_{3/2}+3m_{3/2})\over 20r_1}\right]\nonumber\\
&&-{3\over 5\sqrt{10}}+{\cal O}(s^{1/2}).
\label{2.19}
\end{eqnarray}

\noindent The first term in Eq.~(\ref{2.19}) is proportional to
$\Delta$ and therefore vanishes when there is no phase transition.  The
second term is a pure number that is independent of the parameters of
the phase transition. The second term is therefore the curvature of
the mass-radius curve just below the phase transition point.  The
first term includes a part proportional to $1/s^{1/2}$ which diverges
at the critical model.  This infinite curvature causes the mass-radius
curve to bend sharply in the neighborhood of the critical point.

The analysis here shows that the mass-radius curves of stellar models
with first-order phase transitions have infinite accelerations at the
critical model where the high density phase first appears.  This
acceleration causes these curves to bend sharply, appearing on large
scales to have a discontinuous slope at this point.  Analytical
formulae, Eqs.~(\ref{2.8})--(\ref{2.9}), have been derived that
describe quantitatively the structures of these curves for models with
simple two-zone uniform-density equations of state as illustrated in
Fig.~\ref{fig4}.  These formulae also account in a qualitative way for
the behavior of the mass-radius curves of more realistic equations of
state, as illustrated in Figs.~\ref{fig2}--\ref{fig3}.  The
quantitative description of the more realistic mass-radius curves (as
would be needed to analyze the measured masses and radii of real
neutron stars) requires the derivation of the analogs of
Eqs.~(\ref{2.8})--(\ref{2.9}) for a general equation of state.  This
generalization is not a simple extension of the analysis presented
here, and is deferred to a future investigation.


\acknowledgments

I thank B. Schmidt, J. Friedman, J. Ipser, J. Isenberg and
A. K. M. Masood-ul-Alam for helpful conversations concerning this
work.  I also thank B. Schutz and the Max Plank Institut f\"ur
Gravitationsphysik (Albert Einstein Institut), Potsdam for their
hospitality during a visit in which a portion of this work was
completed.  This research was supported by grants PHY-9796079 from the
National Science Foundation, and NAG5-4093 from the National
Aeronautics and Space Administration.


\appendix
\section*{Continuity of the Slope of the Mass-Radius Curve} 
\label{App}

This appendix presents the argument that the slope of the mass-radius
curve is continuous even at the point where the phase transition first
sets in (except for the case $\Delta=\Delta_c$). The discussion here
is more complete and somewhat more rigorous than Seidov's original
\cite{seidov}.

This argument can be made a little simpler by introducing a somewhat
unusual representation of the equations of stellar structure.  The
structure of a spherical star in general relativity is usually
expressed in terms of the functions $m(r,p_c)$, $p(r,p_c)$ and
$\rho(r,p_c)$ satisfying Eqs.~(\ref{1.1})--(\ref{1.2}). The analysis
here is concerned with how these functions behave as $p_c$ varies near
the point $p_t$ where a phase transition occurs.  Unfortunately the
functions $m(r,p_c)$, $p(r,p_c)$, and especially $\rho(r,p_c)$ are not
smooth when a phase transition is present.  The density function
$\rho(r,p_c)$ in particular is discontinuous at the boundary of the
inner core of high density phase material, and the position of this
discontinuity changes as $p_c$ is varied.  Thus, approximate
expressions for these functions in terms of power series expansions
near the critical point (as needed in Seidov's analysis) are somewhat
awkward.  This difficulty is simplified by considering the structure
of the star in terms of the ``inverses'' of these functions
\cite{lindblom}: e.g. $m(p,p_c)$ and $r(p,p_c)$.  Since the pressure
is a monotonic function of the radius in these models, this inversion
is always possible.  These functions are smooth in their dependence on
$p$ for fixed $p_c$, and so it is more straightforward to approximate
them with power series expansions.

It is also useful to introduce a slightly different set of basic
variables to describe the structures of stars instead of the usual
$p$, $m$, and $r$.  It is preferable to use the thermodynamic enthalpy
function

\begin{equation}
h(p)=\int_0^p {dp\over \rho + p},\label{A1}
\end{equation}

\noindent in place of the pressure as the independent variable in this
representation of the problem, because it makes the differential
equations non-singular at the surface of the star.  Similarly, it is
somewhat preferable to use the functions $u=r^2$ and $v=m/r$ as
dependent variables because they are smoother functions of $h$ near
the centers of the stars.  The straightforward translations of the
standard structure equations (\ref{1.1})--(\ref{1.2}) into this new
set of variables gives

\begin{equation}
{du\over dh} = -{2u(1-2v)\over 4\pi u p(h)+v}
\equiv U(u,v,h),\label{A2}
\end{equation}

\begin{equation}
{dv\over dh} = -(1-2v){4\pi u \rho(h) - v\over 4\pi u p(h) + v}
\equiv V(u,v,h).\label{A3}
\end{equation}
 
\noindent In these Eqs. (\ref{A2})--(\ref{A3}) the functions $p(h)$
and $\rho(h)$ are determined from the chosen equation of state $\rho =
\rho(p)$ and Eq.~(\ref{A1}).  They are therefore explicitly known
functions once a particular equation of state has been selected.  This
version of the equations has several nice features.  First, the use of
$h$ as the independent variable makes the domain where the solution is
defined, $[0,h_c]$ where $h_c$ is the value of $h$ at the center of
the star, known before the solution is found rather than after.
Second, the total radius of the star is determined simply by
evaluating the function $u$ at the surface of the star $h=0$, instead
of solving the usual surface equation $p(R)=0$.  Third, the use of $h$
as independent variable makes the equations non-singular at the
surface of the star.  And fourth, the use of $u$ and $v$ as dependent
variables make the solutions near $h=h_c$ smoother than the
usual functions $m$ and $r$.

Consider the one-parameter family of solutions to these equations
constructed from a single equation of state: $u(h,\lambda)$ and
$v(h,\lambda)$, where $\lambda$ is the parameter that distinguishes the
individual members of the family.  Each member of this family satisfies
the usual boundary conditions, both at the center of the star $h=h_c$,

\begin{equation}
u[h_c(\lambda),\lambda]=v[h_c(\lambda),\lambda]=0,\label{A4}
\end{equation}

\noindent and at the surface of the star $h=0$,

\begin{equation}
u(0,\lambda) = R^2(\lambda),\label{A5}
\end{equation}

\begin{equation}
v(0,\lambda) = {M(\lambda)\over R(\lambda)},\label{A6}
\end{equation}

\noindent where $M(\lambda)$ is the total mass and $R(\lambda)$ is the
total radius of the model with parameter $\lambda$.  The choice of the
particular parameterization is arbitrary; however, it is convenient to
insist that each member of the family have a unique central pressure
$p_c$ and a unique central value of the enthalpy $h_c$.  Thus, either
of these quantities could be used as the parameter $\lambda$.

Near the centers of these stars, the solutions to the structure equations
can be given analytically as power series expansions.  When the equation
of state is smooth [i.e., when $\rho(h)$ and $p(h)$ are smooth functions]
then $u(h,\lambda)$ and $v(h,\lambda)$ have the expansions:

\begin{equation}
u(h,\lambda)= {3(h_c-h)\over 2\pi(\rho_c+3p_c)}
+{\cal O}(h_c-h)^2,\label{A7}
\end{equation}

\begin{equation}
v(h,\lambda) = {2\rho_c(h_c - h)\over \rho_c+3p_c}
+{\cal O}(h_c-h)^2.\label{A8}
\end{equation}

\noindent The right sides of Eqs.~(\ref{A7})--(\ref{A8}) depend on
$\lambda$ implicitly.  The central value of $h$ depends on which
member of the one-parameter family is being considered, thus $h_c =
h_c(\lambda)$.  The choice of parameterization is arbitrary however.
Thus $h_c(\lambda)$ is an arbitrarily monotonic function.  The
quantities $\rho_c$ and $p_c$ also depend on $\lambda$ in the obvious
ways: $\rho_c =\rho[h_c(\lambda)]$, etc.

Next consider the situation where the equation of state is smooth,
except at a certain phase transition point $h=h_{t}$.  Assume
that the density has a finite discontinuity at this point:

\begin{equation}
\label{A8.5}
\rhomin\equiv\lim_{h\uparrow h_{t}}\rho(h) <
\lim_{h\downarrow h_{t}}\rho(h) \equiv \rhoplus.
\end{equation}

\noindent This is simply the restatement of Eq.~(\ref{1.3}) in terms
of $h$ instead of $p$.  The pressure function $p(h)$ is $C^{\,0}$ at
this point as a consequence of Eq.~(\ref{A1}), but it has a finite
discontinuity in its first derivative there.  One of the important
facts that is needed in this analysis is the continuity of the
functions $u$ and $v$.  The functions $u(h,\lambda)$ and
$v(h,\lambda)$ are $C^{\,0}$ functions of $h$ for fixed $\lambda$ and
$C^{\,0}$ functions of $\lambda$ for fixed values of $h$.  These
continuity conditions are reasonably easy to establish.  First
consider the continuity of $u(h,\lambda)$ and $v(h,\lambda)$ as
functions of $h$ for fixed $\lambda$.  The right sides of
Eqs.~(\ref{A2})--(\ref{A3}) are smooth functions of $u$ and $v$ (for
$u>0$ and $v>0$) and are continuous functions of $h$ except when
$h=h_t$.  When $h=h_t$ the right side of Eq.~(\ref{A3}) has a finite
discontinuity as described in Eq.~(\ref{A8.5}).  If $u$ and $v$ were
discontinuous for some value of $h$, then the left sides of
Eqs.~(\ref{A2})--(\ref{A3}) would be singular there.  But the right
sides are finite for $h<h_c(\lambda)$, so $u(h,\lambda)$ and
$v(h,\lambda)$ must be continuous for all $h<h_c(\lambda)$ for fixed
$\lambda$.

Next consider the continuity of $u(h,\lambda)$ and $v(h,\lambda)$ as
functions of $\lambda$ for fixed $h$.  Assume that the parameter
$\lambda$ is chosen so that $h_c(\lambda)$ is smooth and monotonically
increasing.  Let $\lambda_t$ denote the critical value of the
parameter for which $h_c(\lambda_t)=h_t$.  For $\lambda<\lambda_t$ the
expansions in Eqs.~(\ref{A7})--(\ref{A8}) show that $u(h,\lambda)$ and
$v(h,\lambda)$ are continuous in $\lambda$ at least in a small
neighborhood of the center of the star where $h=h_c$.  The
differential Eqs. (\ref{A2})--(\ref{A3}) are non-singular outside of
this neighborhood.  The standard theorems \cite{codd-levin} insure
that the solutions to such non-singular equations depend continuously
on their boundary values.  These boundary values as determined by
Eqs.~(\ref{A7})--(\ref{A8}) can be applied a small distance away from
the singular point $h=h_c$.  Thus, $u(h,\lambda)$ and $v(h,\lambda)$
are continuous functions of $\lambda$ for fixed $h$, at least for
$\lambda<\lambda_t$.  When $\lambda>\lambda_t$ a similar argument
insures the continuity of $u(h,\lambda)$ and $v(h,\lambda)$ in the
cores of the stars where $h\geq h_t$.  Thus $u(h_t,\lambda)$ and
$v(h_t,\lambda)$ are continuous functions of $\lambda$ for
$\lambda>\lambda_t$.  These functions can now be considered as the
boundary values for $u(h,\lambda)$ and $v(h,\lambda)$ in the domain
$h<h_t$.  In this domain the standard theorems again apply, so the
continuity of the boundary values [i.e., $u(h_t,\lambda)$ and
$v(h_t,\lambda)$] guarantee the continuity of $u(h,\lambda)$ and
$v(h,\lambda)$ as functions of $\lambda$ for fixed $h$.  The only
troublesome point is at $\lambda=\lambda_t$.

Consider the stellar models with $\lambda$ just above the critical
point.  These models consist of a very small central core of material
of the higher-density phase, $\rho\geq\rhoplus$, and the vast majority of
the material in the lower-density phase.  In the limit
$\lambda\downarrow\lambda_t$ the size and mass of this central core of
material goes to zero.  This limit can be seen analytically in the
expansions given in Eqs.~(\ref{A7})--(\ref{A8}).  In this limit what
remains is a star composed entirely of matter in the lower-density
phase, except for the single point at the center of the star.  At this
single central point the material remains in the higher-density phase.
But, the matter at this single point does not effect the structure of
the star at all.  The solutions to the structure
Eqs.~(\ref{A2})--(\ref{A3}) are not changed if the equation of state
is changed only at a single value of $h$.  Thus the function
$u_\downarrow(h)=\lim_{\lambda\downarrow\lambda_t}u(h,\lambda)$ is
identical to the function that describes a stellar model consisting
entirely of lower density material with $h_c=h_t$:
$u_\uparrow=\lim_{\lambda\uparrow\lambda_t}u(h,\lambda)$.  A similar
argument applies to $v(h,\lambda)$.  Thus the functions $u(h,\lambda)$
and $v(h,\lambda)$ are continuous functions of $\lambda$ even for
$\lambda=\lambda_t$.

In order to understand the structure of the mass-radius curve in stars
having a first-order phase transition, the structure of stars having a
very small central core of the high density phase material must be
analyzed in some detail.  The central cores of such models are
described by the series solutions given in
Eqs.~(\ref{A7})--(\ref{A8}):

\begin{equation}
\uplus(h,\lambda)= {3(h_c-h)\over 2\pi(\rho_c+3p_c)}
+{\cal O}(h_c-h)^2,\label{A15}
\end{equation}

\begin{equation}
\vplus(h,\lambda)= {2\rho_c(h_c-h)\over \rho_c+3p_c}
+{\cal O}(h_c-h)^2,\label{A16}
\end{equation}

\noindent for $h_t<h<h_c(\lambda)$.  The outer envelopes of these stars
are composed of material from the lower density phase.  In the stars of
interest here---those with only very small cores of high density
material---the structure of the outer envelope is nearly identical to
the structure of a star composed entirely of low density phase material.
Thus the inner region of this outer envelope may be approximated as

\begin{eqnarray}
\umin(h,\lambda)
&= &{3(h_c-h)\over 2\pi(\rhomin+3p_c)}+\delta u(h)\,(h_c-h_t)
\nonumber\\&&+{\cal O}(h_c-h)^2,\label{A17}
\end{eqnarray}

\begin{eqnarray}
\vmin(h,\lambda)
&= &{2\rhomin(h_c-h)\over 
\rhomin+3p_c}+\delta v(h)\,(h_c-h_t)\nonumber\\
&&+{\cal O}(h_c-h)^2,\label{A18}
\end{eqnarray}

\noindent where $\delta u$ and $\delta v$ are solutions to the
linearized structure equations.  Quite generally, these linearized
structure equations have the form

\begin{equation}
{d\delta u\over dh} = {\partial U\over \partial u}\delta u
                      +{\partial U\over \partial v}\delta v,
\label{A19}
\end{equation}

\begin{equation}
{d\delta v\over dh} = {\partial V\over \partial u}\delta u
                      +{\partial V\over \partial v}\delta v,
\label{A20}
\end{equation}

\noindent where $U(u,v,h)$ and $V(u,v,h)$ are the functions defined in
Eqs.~(\ref{A2})--(\ref{A3}).  For our purposes here it is sufficient
to evaluate these functions using the first order terms in the
expansions for $u$ and $v$ given in Eqs.~(\ref{A7})--(\ref{A8}).  In
this case the functions of interest to us have the forms:

\begin{eqnarray}
{\partial V\over \partial u} 
&=& 2\pi(\rhomin+p_c){\partial U\over \partial u}\nonumber\\
&=&-{2\pi(\rhomin+p_c)\rhomin\over
\rhomin+3p_c} {1\over h_c -h}
+{\cal O}(h_c-h)^0,
\label{A21}
\end{eqnarray}

\begin{eqnarray}
{\partial V\over \partial v} 
&=& 2\pi(\rhomin+p_c){\partial U\over \partial v}\nonumber\\
&=&{3(\rhomin+p_c)\over2(\rhomin
+3p_c)} {1\over h_c -h}
+{\cal O}(h_c-h)^0.
\label{A22}
\end{eqnarray}

\noindent The resulting form of Eqs.~(\ref{A19})--(\ref{A20}) can be
integrated analytically.  The general solution is

\begin{equation}
\delta u(h) = A + {B\over (h_c -h)^{1/2}},\label{A23}
\end{equation}

\begin{equation}
\delta v(h) = {4\pi\rhomin\over 3} A + 
{2\pi(\rhomin+p_c)\over (h_c -h)^{1/2}}B,\label{A24}
\end{equation}

\noindent where $A$ and $B$ are arbitrary constants.  The values of
these constants are determined by demanding continuity at $h=h_t$ of the
functions describing the inner core, $\uplus$ and $\vplus$, with the
functions describing the outer envelope, $\umin$ and $\vmin$.  These
continuity conditions are satisfied for the following values of $A$
and $B$:

\begin{equation}
A =-{9(\rho_c-\rhomin)\over 
2\pi(\rho_c+3p_c)(\rhomin+3p_c)},
\label{A25}
\end{equation}

\begin{equation}
B={3(\rho_c-\rhomin)(h_c-h_t)^{1/2}
\over \pi(\rho_c+3p_c)(\rhomin+3p_c)}.
\label{A26}
\end{equation}

\noindent The complete expressions then for the inner portions of the
structure functions in the low-density envelope of the star are

\begin{eqnarray}
\umin(h,\lambda) &=& {3(h_c-h)\over 2\pi(\rhomin+3p_c)}
-{9(\rho_c-\rhomin)(h_c-h_t)\over2\pi(\rho_c+3p_c)(\rhomin+3p_c)}\nonumber\\
&\times&\Biggl[1-{2\over3}\biggl({h_c-h_t\over h_c-h}\biggr)^{1/2}\Biggr]
+{\cal O}(h_c-h)^2,\label{A27} 
\end{eqnarray}

\begin{eqnarray}
\vmin(h,\lambda)& =& {2\rhomin(h_c-h)\over
\rhomin+3p_c}
-{6(\rho_c-\rhomin)(h_c-h_t)\over(\rho_c+3p_c)(\rhomin+3p_c)}\nonumber\\
&&\times\Biggl[\rhomin -(\rhomin+p_c)\biggl({h_c-h_t\over
h_c-h}\biggr)^{1/2}\Biggr]\nonumber\\
&& +{\cal O}(h_c-h)^2.\label{A28}
\end{eqnarray}

\noindent The match of $\umin$ to the inner-core function $\uplus$ is
$C^{\,1}$ at $h=h_t$, as required by Eq.~(\ref{A2}).  The match of
$\vmin$ to $\vplus$ is $C^{\,0}$ at $h=h_t$.  The slope of $\vmin$
differs from that of $\vplus$ at $h=h_t$ by the amount required by
Eq.~(\ref{A3}).

Now consider the region in these models where $|h_c-h_t|\ll |h_c-h|\ll
1$.  The approximate expressions given in
Eqs.~(\ref{A27})--(\ref{A28}) are valid for these models in this
region.  In addition the terms in these expressions proportional to
$(h_c-h_t)^{1/2}$ can be neglected in this region as well.  Next,
evaluate the derivatives $\delta \udown=\partial u/\partial\lambda$,
etc. in this region for these models having a very small core of
high-density phase material:

\begin{eqnarray} \delta \udown(h,\lambda)&=&{\partial\,
\umin\over\partial \lambda} = {3\bigl[-2\rho_c+3(\rhomin+p_{\,c})\bigr]\over
2\pi(\rhomin+3p_{\,c})(\rho_c+3p_{\,c})}{dh_c\over d\lambda}\nonumber\\
&& +{\cal O}(h_c-h)+{\cal O}\biggl({h_c-h_t\over h_c-h}\biggr)^{1/2},
\label{A29}
\end{eqnarray}

\begin{eqnarray}
\delta \vdown(h,\lambda)&=&{\partial\,
\vmin\over\partial \lambda} =
{2\rhomin\bigl[-2\rho_c+3(\rhomin+p_{\,c})\bigr]\over
(\rhomin+3p_{\,c})(\rho_c+3p_{\,c})}{dh_c\over d\lambda}\nonumber\\
&& +{\cal O}(h_c-h)+{\cal O}\biggl({h_c-h_t\over h_c-h}\biggr)^{1/2}.
\label{A30}
\end{eqnarray}

\noindent These expressions can now be compared with those that
pertain to stars having no material at all in the high-density phase.
Thus, define $\delta\uup=\partial u/\partial \lambda$, etc. for the
models with no high density phase material at all using the expansions
in Eqs.~(\ref{A7})--(\ref{A8}) that are valid throughout the inner
regions of these models:

\begin{equation}
\delta\uup(h,\lambda) = {\partial\,u\over\partial\lambda}
= {3\over 2\pi(\rhomin+3p_c)}{dh_c\over d\lambda}+{\cal
O}(h_c-h),\label{A31}
\end{equation}

\begin{equation}
\delta\vup(h,\lambda) = {\partial\,v\over\partial\lambda}
= {2\rhomin\over \rhomin+3p_c}{dh_c\over d\lambda}+{\cal
O}(h_c-h).\label{A32}
\end{equation}

\noindent These expressions illustrate that the derivatives $\delta u$
and $\delta v$ are not continuous functions of $\lambda$ near the
critical model with $\lambda=\lambda_t$.  However, the discontinuity
is of a very special type.  These expressions for $(\delta \udown,
\delta \vdown)$, are related to those for $(\delta \uup,\delta
\vup)$ in the following simple way,

\begin{equation}
\left({\delta\udown \atop \delta\vdown}\right)=
2{\Delta_c-\Delta\over 2\Delta_c+\Delta}
\left({\delta\uup \atop \delta\vup}\right),\label{A33}
\end{equation}

\noindent where $\Delta$ and $\Delta_c$ are defined in
Eqs.~(\ref{1.4}) and (\ref{1.5}) respectively.  Equation (\ref{A33})
is exact in the limit that $\lambda\rightarrow \lambda_t$ from above
and below respectively, and when the functions are evaluated at the
center of the star $h=h_t$.  The derivatives $\delta u$ and $\delta v$
satisfy the linear differential Eqs.~(\ref{A19})--(\ref{A20}).
Further, the continuity of $u$ and $v$ as functions of $\lambda$ at
the point $\lambda=\lambda_t$, implies that both $(\delta\uup,
\delta\vup)$ and $(\delta\udown,\delta\vdown)$ satisfy the
{\it same} differential equation at $\lambda=\lambda_t$.  (This fact
is the reason that it was necessary to establish the continuity of $u$
and $v$ in some detail above.)  Thus, it follows that the functions
$(\delta \uup,\delta \vup)$ are proportional to $(\delta \udown,
\delta \vdown)$ throughout the critical model with
$\lambda=\lambda_t$ since they are proportional to one another in a
neighborhood of $h=h_t$.

At the surface of the star, $h=0$, the derivatives $\delta u$ and
$\delta v$ are related to the total mass and radius of the star as a
consequence of Eqs.~(\ref{A5})--(\ref{A6}). In particular these
functions must satisfy:

\begin{equation} \delta u(0,\lambda) 
= 2 R(\lambda) {dR(\lambda)\over d\lambda},
\label{A34}
\end{equation}

\begin{equation}
\delta v(0,\lambda) = {1\over R(\lambda)}{dM(\lambda)\over d\lambda}
   - {M(\lambda)\over R^2(\lambda)}{dR(\lambda)\over d\lambda}.
\label{A35}
\end{equation}

\noindent The functions $(\delta\udown,\delta\vdown)$ (evaluated
for models just above the critical one) are proportional to
$(\delta\uup,\delta \vup)$ (evaluated for models just below the
critical one) throughout the star.  Thus, the surface values of these
functions are proportional as well.  This implies in particular that
the tangent vectors to the mass radius curve evaluated above and below
the critical model are related by:

\begin{equation}
\left({dM_{\downarrow}\over d\lambda}\atop
               {dR_{\downarrow}\over d\lambda}\right)=
2{\Delta_c-\Delta\over 2\Delta_c+\Delta}
\left({dM_{\uparrow}\over d\lambda}\atop
               {dR_{\uparrow}\over d\lambda}\right).
\label{A36}
\end{equation}

\noindent This expression has several interesting consequences.
First, it shows that the tangent vector to the mass-radius curve is
discontinuous at the critical model whenever one parameterizes the
curve in a way that makes $h_c(\lambda)$ smooth.  Second, this
expression shows that the mass-radius curve in fact reverses direction
at the critical model if the phase transition is sufficiently severe
so that $\Delta>\Delta_c$.  Third and finally, Eq.~(\ref{A36}) implies
that the slope of the mass-radius curve, $dM/dR$, is continuous even
at the critical model:

\begin{equation}
{dM\over dR}=
{dM_{\downarrow}\over d\lambda}
\left({dR_{\downarrow}\over d\lambda}\right)^{-1}
={dM_{\uparrow}\over d\lambda}
\left({dR_{\uparrow}\over d\lambda}\right)^{-1}.
\label{A37}
\end{equation}

\noindent Continuity of the slope pertains even if the curve has a
cusp and reverses direction at the critical model, unless
$\Delta=\Delta_c$.  In this special case Eq.~(\ref{A36}) merely
implies that $\lambda$ [chosen so that $h_c(\lambda)$ is smooth] is
not a good affine parameter for the mass-radius curve.  A higher order
analysis is needed to understand the differentiability of the curve in
this special case.


\end{document}